\documentclass[useAMS,usenatbib,usegraphicx]{mn2e}

\usepackage{aas_macros}
\usepackage{multirow}

\usepackage{color} 

\usepackage{lineno}

\def\gx{GX~339$-$4}

\def\1e{1E~1740.7$-$2942}

\def\xteonze{XTE~J1118$+$480}

\def\vq{V404~Cyg}
\def\a0{A~0620$-$00}
\def\1h{H~1743$-$322}
\def\4u{4U~1755$-$33}

\def\ergs{\hbox{erg s$^{-1}$}}

\def\um{{$\mu$m}}                                       
                                       
  \def\her{{\it Herschel}}

\title[Formation of compact jets]{Formation of the compact jets in the black hole  \gx .   }
\author[S. Corbel et al.]{S. Corbel$^{1}$\thanks{E-mail:stephane.corbel@cea.fr}, H. Aussel$^{1}$, J. W. Broderick$^{2}$, P. Chanial$^{1}$,   M. Coriat$^{2}$,   A. J. Maury$^{3}$,
\newauthor 
 M. Buxton$^{4}$,  J. A. Tomsick$^{5}$, A. Tzioumis$^{6}$, S.  Markoff$^{7}$,   J. Rodriguez,$^{1}$ C. Bailyn$^{4}$,    
\newauthor 
C. Brocksopp$^{8}$, R.  Fender$^{2}$, P.O. Petrucci$^{9}$,  M. Cadolle-Bel$^{10}$, D. Calvelo$^{2}$, L. Harvey-Smith$^{6}$
  \\
$^{1}$Laboratoire AIM (CEA/IRFU - CNRS/INSU - Universit\'e Paris Diderot), CEA DSM/IRFU/SAp, F-91191 Gif-sur-Yvette, France \\
$^{2}$School of Physics and Astronomy, University of Southampton, Highfield, Southampton, SO17 1BJ, UK \\
$^{3}$ESO, Karl Schwarzschild Strasse 2, 85748, Garching bei Munchen, Germany\\
$^{4}$Astronomy Department, Yale University, P.O. Box 208101, New Haven, CT 06520-8101, USA\\
$^{5}$Space Sciences Laboratory, 7 Gauss Way, University of California, Berkeley, CA 94720-7450, USA\\
$^{6}$CSIRO Astronomy \& Space Science, Australia Telescope National Facility, PO Box 76, Epping NSW 1710, Australia\\
$^{7}$Astronomical Institute 'Anton Pannekoek', University of Amsterdam, PO Box 94249, 1090 GE Amsterdam, the Netherlands \\
$^{8}$Mullard Space Science Laboratory, University College London, Holmbury St Mary, Dorking, Surrey RH5 6NT, UK\\
$^{9}$UJF-Grenoble 1/CNRS-INSU, Institut de Plan\'etologie et d'Astrophysique de Grenoble (IPAG) UMR 5274, Grenoble F-38041, France\\
$^{10}$ESAC/ISOC, P.O. Box 78, 28691 Villanueva de la Ca\~nada, Spain.
}
\begin{document}

\date{Accepted XXXX. Received XXXX; in original form 2012 October XX}

\pagerange{\pageref{firstpage}--\pageref{lastpage}} \pubyear{2002}

\maketitle

\label{firstpage}

\begin{abstract}
Galactic black hole binaries produce powerful outflows with emit over almost the entire electromagnetic spectrum. 
Here, we report the first detection with the \her\  observatory  of a variable  far-infrared source associated with the compact jets of the black hole transient \gx\ during the decay of its recent 
2010-2011 outburst,   after  the transition to the  hard state.  We also outline the results of   very sensitive radio observations conducted with the Australia Telescope Compact Array, 
along with a series of near-infrared,  optical (OIR) and X-ray observations, allowing  for the first time  the re-ignition  of the compact jets to be observed over a wide range of  wavelengths. 
The compact jets first turn on at radio frequencies with an optically thin spectrum that later evolves to optically thick  synchrotron emission. An OIR reflare is observed about ten days after 
the onset of radio and hard X-ray emission, likely reflecting the necessary time to build up enough density,  as well as to have  acceleration (e.g. through shocks) along an extended region in the jets. 
The \her\ measurements are  consistent with an extrapolation of the radio inverted power-law spectrum, but they highlight a  more complex radio to OIR spectral energy distribution for the jets. 
\end{abstract}


\section{Introduction}

\begin{figure*}
\begin{tabular}{cccc}
\includegraphics[width=0.2\textwidth]{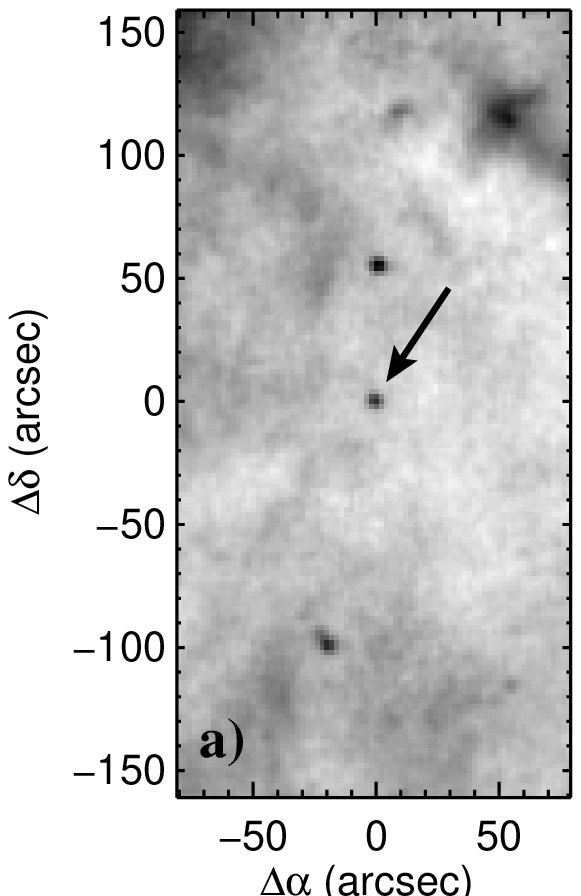} & \includegraphics[width=0.2\textwidth]{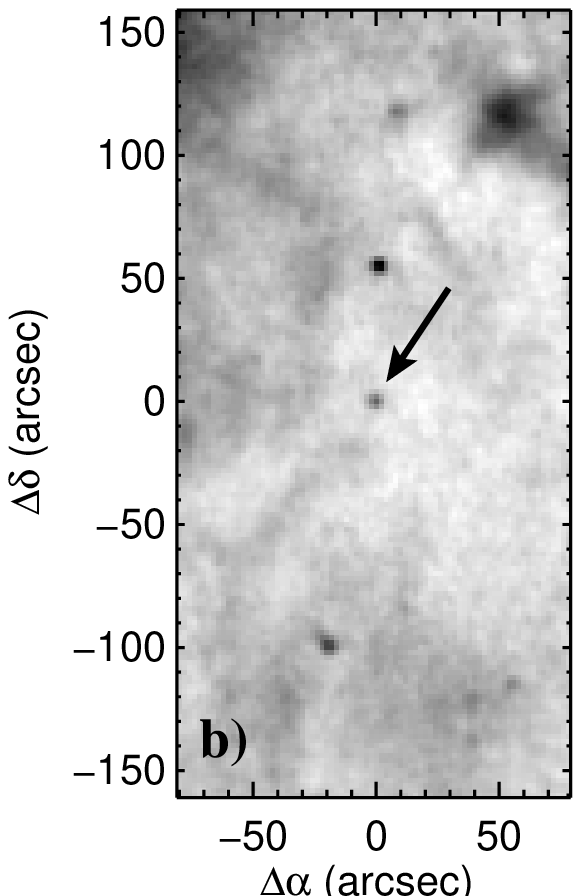} & \includegraphics[width=0.2\textwidth]{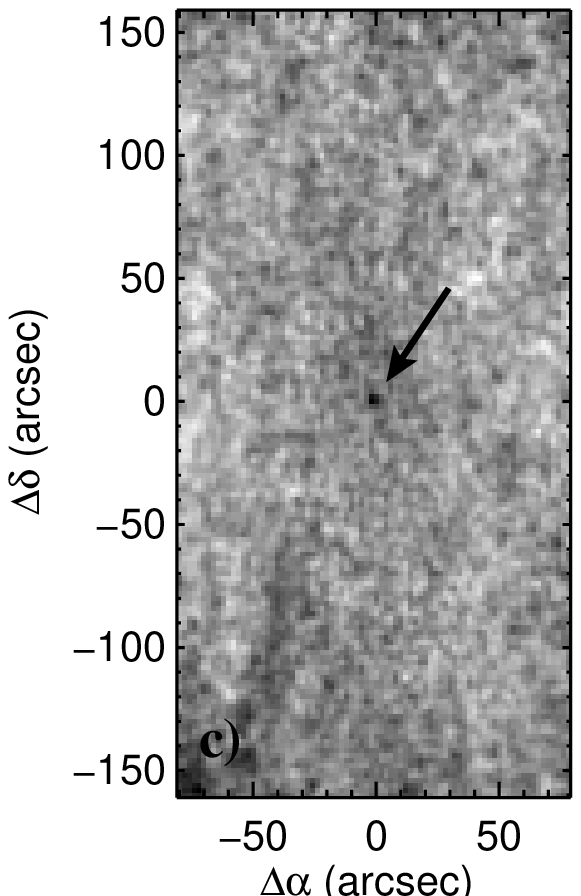} & \includegraphics[width=0.2\textwidth]{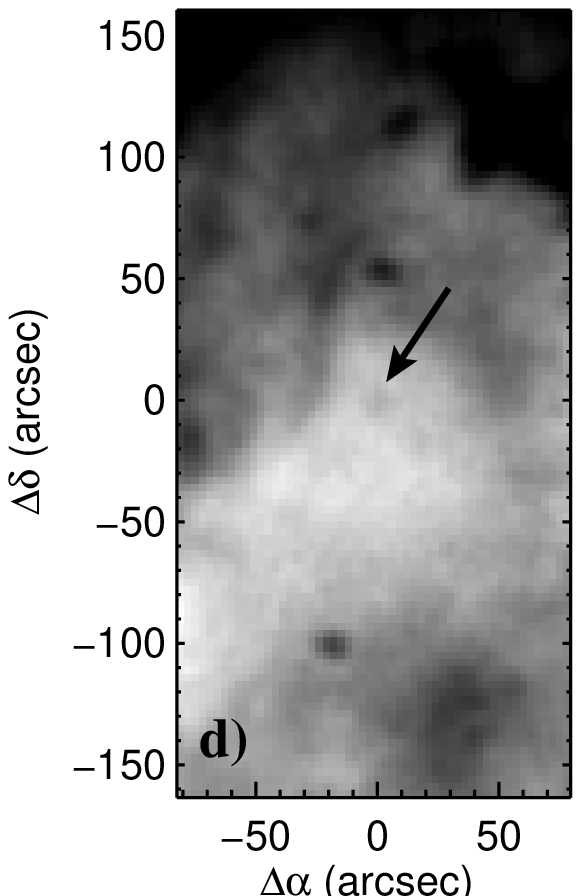} \\
\end{tabular}
\caption{\her\  maps of \gx :  at 70~$\mu$m on Feb. 25 (a) and  Mar. 6  (b),  their difference in (c), and  at 160~$\mu$m (2 epochs) in (d).   }

\label{fig_hersc}
\end{figure*}

During  outbursts of X-ray binaries,  an enormous amount of accretion energy is harnessed to produce powerful jets \citep[e.g.][]{Fender06b}.
For a very long time,  those jets were only detected in radio \citep[e.g.][]{Hjellming75,Mirabel92}. However, the recent decade
 demonstrated that relativistic jets from black holes  (hereafter BH) and neutron stars are multi-wavelength emitters with emission
  even up to $\gamma$ rays   \citep{Laurent11, Corbel12}. Powerful self-absorbed compact jets \citep{Fender01}  are observed  in 
the hard state of  BH binaries, and are reminiscent of  those found in active galatic nuclei. 
Models of  compact jets  \citep{Blandford79,Falcke95} predict that their spectral energy distributions (SED) should be   
 a  flat or slightly inverted power-law  between the radio and a turnover frequency estimated  in the mid to near-infrared (NIR) for Galactic BHs \citep{Corbel02a,Gandhi11, Russell12b}. 
Such NIR synchrotron excess emission from the jets has now been detected in a growing number of BH transients (e.g. \citealt{Jain01, Buxton04, Buxton12, Dincer12, Russell12}). 
However, there is almost an unexplored four decades frequency gap  between the radio and the NIR  domains. Only very few millimeter and 
mid-infrared (IR) observations  \citep[e.g.][]{Fender01b,Migliari07,Gallo07,Rahoui11,Gandhi11} have been conducted in this range and usually  with  very limited sensitivity.
New theoretical calculations highlight  the possibility of a more complicated spectral  shape \citep{Peer09}, especially in the  
 far to NIR range. 
Strong variability  can be observed in the NIR on  short  timescale ($\sim$ seconds) allowing one to probe physical changes at the base of the jet.
These variations seem to  vanish  on longer timescales   (above a few hundreds seconds,  \citealt[][]{Casella10, Gandhi11,Rahoui12}).

This points to  the importance of sensitive sub-millimeter to far-IR observations that   can only bridge the gap between the 
radio and the NIR ranges. In this Letter, we report the first  far-IR observations with the ESA  {\it Herschel} satellite of the recurrent 
binary \gx\ during the decay of its  recent outburst in 2011. \gx\ is believed to harbour a BH with a mass $>$ 7 ~$M_{\odot}$ \citep{Hynes03b,MunozDarias08} in a system 
with  a low mass star companion, and located at a likely distance of 8 kpc \citep{Zdziarski04}.
In Section 2, we describe the {\it Herschel} observations along with complementary radio, NIR, optical and X-ray observations. The detailed multi-wavelength 
lightcurves, as well as the far-IR detections,  are also presented and  these results are discussed in Section 3.

\section{Observations and Results}

\subsection{Herschel far-infrared Observations}

Following the transition of \gx\  back to the hard state during the decay of its 2010-11 outburst, we triggered two DDT (Director Discretionary Time) observations with the  {\it Herschel} satellite\footnote{\her\ is an 
ESA space observatory with science instruments provided by  European-led Principal Investigator consortia and with important participation from NASA.}  \citep{Pilbratt10}.
\gx\ was observed with the PACS instrument \citep{Poglitsch10}  at 70~$\mu$m and 160~$\mu$m on 2011, February 25  and March 6 (Table~\ref{tab_obs}). 
We used a sequence of two so-called ``mini-scan map" mode, providing an area of 50" in diameter with homogenous coverage for a total on-source time of 1152 s. 
The data were processed with {\tt HCSS} version 6.0.1106   
and we reconstructed the maps  with the {\tt Tamasis} software (Chanial et al. in prep)\footnote{see {\tt http://pchanial.github.com/tamasis-pacs/}}.   
As \gx\  sits in the Galactic plane in the midst of a strong cirrus background,  we paid a special attention in the map reconstruction to preserve the structure of the background.
The resulting flux density  maps at 70 $\mu$m are shown in Fig.~\ref{fig_hersc} (a, b). A point source is clearly detected at the center of each  map, as well as some extended emission across the  field. 
A single point source is visible in the difference image (Fig.~\ref{fig_hersc}c) at the center, clearly indicating the detection at 70~$\mu$m of a variable source
 at the location of \gx . All the other sources and structure have been subtracted out. 
 In order to choose the optimum apertures for source and sky extraction, calibrate our photometry and check the impact of the varying cirrus background, 
we have  run a set of simulations, where point sources were added to the bolometer timelines by applying our observation model to the  
observed PACS point spread function of \citet{Lutz12}. We reconstructed the maps and performed aperture photometry on the simulated sources and \gx , 
 with various aperture size and sky annuli. We selected the combination giving the highest signal to noise for point sources (Table~\ref{tab_obs}). The cirrus confusion noise adds 
0.77 mJy at 70~$\mu$m and 4.5 mJy at 160~$\mu$m in quadrature to the point source sensitivity. As a result, we only obtain a 2.9 $\sigma$ detection at 160~$\mu$m by  combining  the two 
observations,   while  \gx\ is detected at high significance at 70~$\mu$m in each  epoch.

\begin{table}
\begin{tabular}{|c|c|c|c|}
\hline
Obs. ID & Date  & $f_{\nu}$ (70 $\mu$m) &  $f_{\nu}$ (160 $\mu$m) \\
             &          &  (mJy)  & ( mJy) \\
\hline
1342215720 & 2011-02-25  & \multirow{2}{*}{21.0 $\pm$ 1.0} & \multirow{4}{*}{13.2 $\pm$ 4.6}\\
1342215721 & 2011-02-25  & \\
\cline{1-3}
1342216156 & 2011-03-06  &  \multirow{2}{*}{13.9 $\pm$ 1.0} & \\
1342216157 & 2011-03-06   & \\
\hline
\end{tabular}
\caption{\her\ observations log of \gx }
\label{tab_obs}
\end{table}

\subsection{Radio, OIR and X-ray  observations}

In order to probe the evolution of the SED from the compact jets,  the  {\it Herschel} observations were included in a large campaign of multi-wavelength observations. 
All radio observations were performed with the Australia Telescope Compact Array (ATCA)  located in Australia. The ATCA observations 
were  conducted at 5.5 and 9 GHz (also at 2.2 GHz during the first  {\it Herschel} run) with the  upgraded and  sensitive CABB  back-end (see Table 2 available as online supporting information). 
The PCA and XRT detectors onboard the {\it RXTE} and {\it Swift} satellites provide the  X-ray coverage during the soft to hard state transition.  Further details on the radio and X-ray data analysis can be found in \citet{Corbel12b}.
The optical and near-infrared (herafter OIR) observations were taken with the ANDICAM camera on the SMARTS 1.3~m 
telescope located at Cerro Tololo in Chile (see \citealt[][]{Buxton12}). Images in $H, J, I$ and  $V$ filters are obtained on a near-daily basis when \gx\ is visible in the night sky. 
The  magnitudes have been converted to flux units using a total extinction of  A$_\mathrm{V}$ = 3.7 mag. 

\begin{figure}
\includegraphics[width=88mm]{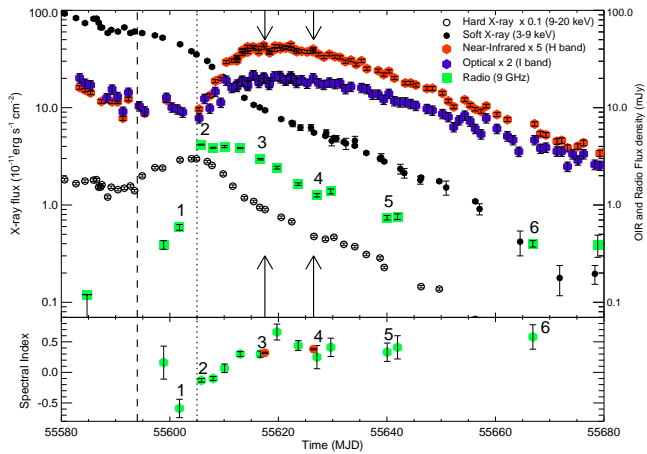}
\caption{ (Top): The lightcurves (with multiplicative factors as indicated in the legend)  of  \gx\ during the decay of the 2010-11 outburst in radio (green), NIR (red), optical (purple), soft X-ray 
(filled black circle) and hard X-ray (open black circle). The arrows highlight the  \her\  observations (observations  \# 3 and 4). 
(Bottom): Evolution of the radio (green hexagon) 5.5 to 9 GHz spectral index (see text) once \gx\ initiated its transition back to the hard state. The radio to far-IR spectral indices
using the {\it Herschel} data) are represented by the red circle.   The dashed line  indicates the transition back to the hard state according to \citet{Dincer12}. The 
vertical dotted line highlights the onset of OIR emission close to the peak of hard X-ray and radio emission. The numbers refer to the individual observations displayed in Fig. \ref{fig_sed}.   }
\label{fig_lc}
\end{figure}

\subsection{The  re-ignition of the  compact jets}

Figure \ref{fig_lc} shows the multi-wavelength lightcurves of \gx\ during the decay of its 2010-11 outburst. The dashed line indicates the increase 
of the hard X-ray emission on MJD 55594 associated with the transition of \gx\  back to the hard state (see  \citealt{Dincer12}). 
Radio emission from  the compact jets is  quenched in the soft state \citep{Fender99}. Due to the increased sensitivity of the ATCA array and denser coverage,  we fully observe, for the first time, 
the recovery of the radio emission during the soft to hard state transition when the compact jets are building up.
Despite sparse initial coverage, the onset of radio emission from \gx\ is clearly observed.  It occurs in conjunction  with the increase of hard X-ray emission (Fig.~ \ref{fig_lc}).  
The peak of radio emission is concomitant with the peak in hard X-rays, as well as with the rise of the OIR emission. However, the
overall shape of the reflares are different with, e.g.,  the hard X-rays appearing  to be more symmetric than the radio peak.  
The peak flux in OIR  is reached $\sim$ 2 weeks after the maximum of radio emission. 
The timescale to reach the peak emission  (assuming a transition on MJD 55594) is much  longer ($\sim$ 17 days)  in OIR compared to hard X-rays ($\sim$ 10 days), and maybe also compared  to radio. 
 We note that the transition may have even started in radio and X-ray as early as  MJD 55578, if we consider  the weak radio detection (see Table 2 online) that is consistent with the leveling off of the  X-ray timing properties \citep{Dincer12}.

If we define the radio flux density, F$_\nu$,  as F$_\nu \propto \nu^\alpha$ (with $\alpha$  the  spectral index), we note  a clear change,
 after  the transition to the hard state (Fig. \ref{fig_lc}), from a negative spectral  index to a saturation around   $\sim$ +0.4. 
This trend corresponds to an evolution from optically thin to optically thick synchrotron emission, associated with the building up of the
  compact jets during the  state transition.
Connecting the radio measurements to the two \her\ observations at 70 \um\ allows  us to precisely characterise the radio to far-IR spectral indices to 0.32 $\pm$ 0.01 
on Feb. 25th and 0.38 $\pm$ 0.01 on Mar 6th, which is fully consistent with the radio estimates (Fig. \ref{fig_lc}). 
The detections  by  \her\ of a variable source at the location of  \gx\   during a hard state further  indicate that this far-IR emission is likely related to the self-absorbed compact jets of \gx .

\section{Discussion}

Our  campaign during the decay of the 2010-11 outburst of \gx\  allowed  us  to obtain for the first time a  significant detection of the compact jets
in the far-IR with \her , as well as to observe  the full recovery of the compact jets  in radio and OIR. 
This enables us to bridge the 4-decade frequency gap between the radio and OIR domains.

\subsection{Formation of the compact jets}\label{sect_form}

Inspection of the evolution of the broadband spectra  (Fig. \ref{fig_sed} ) demonstrates that the compact jets are first detected at radio frequencies as optically thin synchrotron emission (\# 1 and 2). 
The radio spectrum becomes  flat  (Fig. \ref{fig_lc} (bottom)) after the onset of the OIR reflare  
and its spectral index then increases up to a value of the order of +0.40, at a time when the OIR peaks and starts to decay. Such sequence has not been documented before
in as much details (e.g. the 1999 decay of \gx, \citealt{Corbel00}),  especially in light of the  evolution of the radio emission during the soft to hard state transition. 
However, the previous studies that discuss some radio and/or OIR observations \citep[][]{Kalemci05,Kalemci06, Brocksopp05,Fender09,MillerJones11,Dincer12,MillerJones12b} are 
consistent with our new findings. 
The comparative evolution of the radio and OIR lightcurves clearly points to a delay ($\sim$ 2 weeks) between these two domains (see also \citealt{Coriat09}). 
In fact, this is consistent with the reverse trend that is observed during the hard to soft state  transition, with an initial quenching  (by one to two weeks) of the NIR emission \citep{Homan05, Russell08, Coriat09, Buxton12,Yan12}. 

In the model of self-absorbed compact jets \citep{Blandford79,Falcke95} adapted for stellar mass BHs \citep[e.g.][]{Markoff05}, the 
OIR emission  arises from closer to the base of the jets,
whereas the radio emission  comes further down the jets. 
The observed delay between OIR and radio emission during the formation is not consistent with any travel time  within the jets.  
We still do not fully understand the internal processes in jets leading to particle acceleration and ultimately radiation. However, this delay is likely associated with the timescale
for building up turbulences and strong shocks along an extended zone in the magnetized jet plasma, as well as for reaching high enough densities for the flow to become 
optically thick at OIR frequencies.
One can imagine the following scenario to explain the evolution of the broadband spectra. Once the jets are launched, an initial zone of particle acceleration, associated with the first shocks, forms. 
While the density of the jet plasma is still low, this  results in an optically thin synchrotron power-law spectrum peaking in the radio band.
As the density in the jets increases, this leads to a transition to higher optical depth.  
At the same time,  pile-up of the flow or internal shocks \citep{Malzac12} build up and cause particle acceleration to  occur closer to the  base of the jets.  Alternatively, an increase in particle acceleration 
efficiency at the jet base could also boost the energy of the  particles, giving rise gradually to enhanced synchrotron emission up to  OIR.

One would therefore expect the observed behaviour: a transition from a steep to a flat spectrum with an evolution of the peak power from  radio to  NIR.
A reverse scenario (decrease in the particle densities and/or the efficiency of particle acceleration) would also be consistent with 
the  trend (initial OIR  quenching) observed during the hard to soft state transition.  The onset of radio emission can thus be seen as a tracer (better than  OIR) of the jets re-ignition, 
while the broadband evolution 
leading to inverted  synchrotron spectrum peaking in the OIR  
traces the  development of the jets power and their internal structure during outbursts. 

\subsection{On the far- to near-infrared emission}

\begin{figure}
\includegraphics[width=85mm]{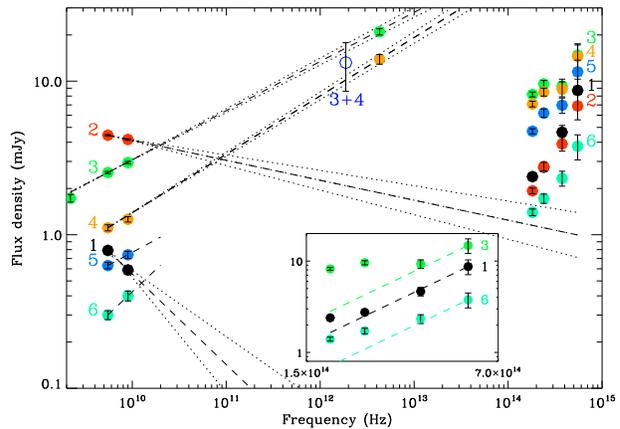}
\caption{Evolution of the radio to OIR broadband spectra during the soft to hard state transition. The numbers (and associated colors) refer to the epochs highlighted in Fig. \ref{fig_lc}. The dashed
lines corresponds to the extrapolation of the radio spectra (with dotted lines to indicate the 1-$\sigma$ error when it is not too large). For the spectra \# 3 and 4 (which includes the \her\ observations), we
use the measured radio to far-IR spectral index. The  open blue circle
is the 160 \um\ detection combining the two \her\ observations.  Inset: Zoom of the OIR spectra \# 1, 3 and 6. The dashed lines correspond to an extrapolation of the V-band with an index of +1.5 (possible for an irradiated  disk, 
\citealt{Hynes05}) for a selection of the spectra. }
\label{fig_sed}
\end{figure}

The very good sensitivity of the \her /PACS instrument and  of the upgraded ATCA demonstrates that the far-IR emission of  \gx\  is clearly and significantly above  a simple  
extrapolation of the radio  to NIR spectrum (see broadband spectral  evolution in Fig. \ref{fig_sed} during the phase of compact jets formation). This is in contradiction with the usual 
assumption of a simple powerlaw \citep[e.g.][]{Markoff05} with a turnover frequency in the mid to  near-IR range. 
Subtracting the thermal component in the OIR  would further increase the far-IR emission relative to the OIR. This is in fact reminiscent of the 850 \um\ (350 GHz) JCMT detection 
of \xteonze\ in 2000  \citep{Fender01b}, which was also  clearly above the radio to OIR inverted spectrum (see their  Fig.~2). 
The radio to sub-millimeter spectral index of  +0.5 for \xteonze\ is furthermore quite  comparable with the value measured during our \her\ observations of \gxÊ (Fig.~\ref{fig_lc}). \vq\ may 
 display occasionally  a similar broadband spectra \citep{Russell12b}. 

These three sources seem to point out  a more complex broadband spectra for the compact jets during some epochs. The optically thick synchrotron emission from the jets manifests itself by an inverted
power-law from the radio up to at least the far-IR (same spectral index, Fig.~\ref{fig_lc}). The turnover frequency  during our two \her\ observations  is likely  
above 5 $\times$ 10$^{12}$ Hz, which is  consistent with previous  estimates.   It is clear from the evolution of the OIR lightcurves (Fig.~\ref{fig_lc}  and also  \citealt{Dincer12}) that most
of the NIR emission during the reflare is not from the intrinsic decaying thermal emission from the accretion disk. Furthermore, this OIR reflare can not be completely  due to  the optically 
thin synchrotron emission from the compact jets, because it displays a rather flat spectra (after subtraction of the thermal emission) according to \citet{Dincer12}.  
This  means  that another emission  process may  dominate  the OIR range (see also \citealt{Rahoui12,Dincer12}).  Even if the OIR reflare is delayed compared to the radio synchrotron emission, this
OIR emission could still be related to the jets via, e.g.,  pre-shock synchrotron  from the thermal particles at the base of the jets \citep{Markoff05}.  The OIR could  also be affected 
by an episode of particle acceleration in the jets in a particularly strong magnetic field \citep{Peer09}. 
Alternatively,  the OIR excess  may be related to synchrotron emission from the hot accretion flow \citep{Veledina11}. 
At first sight, the shape of the OIR spectra (see inset in Fig.  \ref{fig_sed}) could also possibly highlight, 
as a viable alternative mechanism, the role of irradiation of the disk  (Hynes 2005) by the jets, which form during this period.

With optically thick synchrotron emission in the 10$^9$~-5~$\times$~10$^{12}$ Hz  ($\sim$ 70 \um ) range and assuming a radiative efficiency of $\epsilon$=0.05 
[see discussion in \citet{Corbel02a},  or recently in  \citet{Nemmen12} who  suggest that  $\epsilon$ could be as high as 0.15], we
obtain a likely lower limit on the power of compact jets  during  \her\ observations \#1  of $\sim$ 10$^{35}$ \ergs . 
Adding an optically thin jet component (of spectral index  --0.6)  in the 5 $\times$ 10$^{12}$ to 10$^{14}$ Hz  increases the jet power   to $\sim$  10$^{36}$ \ergs .
This corresponds to at least 17 \%  of the 0.1-200 keV  luminosity (6 $\times$ 10$^{36}$ \ergs\ for a distance of 8 kpc), a significant fraction of the accretion power (assuming no correction for bulk relativistic motion). 
This probably implies that the jet spectral break   can not be much higher than the frequency of the \her\ observations.  For a comparison, in the brighter hard state during the rise (with an X-ray luminosity of 1.5 $\times$10$^{38}$ \ergs ),
 \citet{Gandhi11} measured this break at $\sim$ 5 $\times$ 10$^{13}$ Hz.  
A jet frequency break moving to lower frequency is expected as the mass accretion is reduced during the outburst decay \citep{Heinz03,Falcke04,Russell12b}. 
Furthermore,    the \her\ measurements would translate into a magnetic field strength  of  $\sim$ 2 $\times$ $10^3$ G in the acceleration zone 
and a jet cross section radius of $\sim$ $10^{10}$ cm \citep{Chaty11}. These numbers would eventually imply  compact jets with weaker magnetic field and a larger base during 
the decaying hard state compared to the rising hard state observed by \citet{Gandhi11}  (if we take the same assumptions).

To summarize, our new campaign highlights a much more complicated radio to OIR spectral shape than usually assumed, especially around the time of the  transition. However,  
on many occasions, the OIR measurements may still lie on the extrapolation of the radio spectrum  \citep[e.g.][]{Coriat09,Buxton12}.
This reinforces the need for new observations, e.g. with the  ALMA array or with sensitive mid to far IR observations,  to bridge the gap between the radio and the OIR domains and constrain the  evolution of the jets SED and power during BH outbursts. 

\vspace{-0.5cm}
\section*{Acknowledgments}

We thank  G\"oran Pilbratt  for accepting this program under  Director Discretionary  Time and Bruno Altieri for help with the observations. 
This research has received funding from the European Community  (FP7/2007-2013) ITN 215212 ÓBlack Hole Universe" and the ANR "CHAOS" (ANR-12-BS05-0009).
SC would like to thank Phil Edwards for prompt
scheduling of the ATCA observations. The Australia Telescope is funded by the Commonwealth of Australia for operation as a national
Facility managed by CSIRO.
Tamasis was developed with funding of the E.C. FP7 ASTRONET program. 
PACS has been developed by a consortium of institutes led  by MPE (Germany) and including UVIE (Austria); KU Leuven, CSL, IMEC 
(Belgium); CEA, LAM (France); MPIA (Germany);  INAF-IFSI/OAA/OAP/OAT, LENS, SISSA (Italy); IAC (Spain). This development has 
been supported by the funding agencies BMVIT (Austria), ESA-PRODEX (Belgium), CEA/CNES (France), DLR (Germany), ASI/INAF (Italy), and CICYT/MCYT (Spain).
JAT acknowledges partial support from NASA {\em Swift} Guest Observer grant NNX10AK36G and also from
the NASA Astrophysics Data Analysis Program grant NNX11AF84G. 
MMB and CDB are supported by NSF/AST grants 0407063 and 070707 to CDB. 
POP acknowledges financial support from CNES and GDR PCHE. 
This research has made use of data obtained from the High Energy Astrophysics
Science Archive Research Center (HEASARC), provided by NASA's Goddard
Space Flight Center. 

\bibliographystyle{mn2e_fixed}


\end{document}